\documentclass[sigconf]{acmart}
\usepackage{amsfonts}
\usepackage{multirow}
\usepackage{verbatim}
\usepackage[font=small]{caption}
\usepackage{cleveref}
\usepackage{tabularx}
\usepackage{enumitem}
\usepackage{makecell}
\usepackage{siunitx}
\usepackage{xcolor}
\usepackage{threeparttable}

\newcolumntype{Y}{>{\centering\arraybackslash}X}
\usepackage{amsmath,bm}
\usepackage{subcaption}
\usepackage{graphicx}

\usepackage{booktabs}
\usepackage{xcolor}
\usepackage{colortbl}
\usepackage{multirow}
\usepackage{rotating}

\begin{document}

\title{HySpecPro: Scalable Hypergraph Partitioning via Spectral Projection Optimization}

\author{Rongjian Liang}
\email{rliang@nvidia.com}
\affiliation{
  \institution{NVIDIA}
  \city{Santa Clara}
  \country{USA}
}

\author{Zhuo Feng}
\email{zfeng12@stevens.edu}
\affiliation{
  \institution{Stevens Institute of Technology}
  \city{Hoboken}
  \country{USA}
}
\affiliation{
  \institution{NVIDIA}
  \city{Santa Clara}
  \country{USA}
}

\author{Haoxing Ren}
\email{haoxingr@nvidia.com}
\affiliation{
  \institution{NVIDIA}
  \city{Santa Clara}
  \country{USA}
}

\begin{abstract}

Modern VLSI designs comprise tens of billions of components, making scalable hypergraph partitioning critical for parallel and hierarchical optimization. Although multilevel partitioning remains the dominant paradigm, its coarsening stage can distort structural information—especially in hypergraphs with many high-degree hyperedges—leading to increased refinement overhead and limited scalability. Recent approaches incorporate spectral information to guide coarsening, but only in a heuristic manner, without directly optimizing the partitioning objectives. We introduce HySpecPro, a single-level hypergraph partitioner that performs end-to-end optimization in a spectral embedding space. HySpecPro constructs embeddings from a bipartite Laplacian and performs efficient projection-based search, supported by a fully GPU-accelerated implementation. Experiments show that HySpecPro delivers cut quality comparable to state-of-the-art multilevel methods while scaling linearly with the total hyperedge degree.

\end{abstract}

\maketitle

\section{Introduction}
Modern VLSI designs comprise tens of billions of interconnected components, making full-chip optimization computationally prohibitive and often requiring weeks to months to complete. To improve scalability, circuit netlists are typically decomposed into smaller sub-blocks for parallel or hierarchical optimization. Hypergraphs provide a natural representation of netlists by capturing multi-way connectivity among circuit elements. A central task in this context is hypergraph partitioning~\cite{karypis1997multilevel}, which divides vertices into balanced parts while minimizing objectives such as cut size. However, the problem is NP-hard due to its discrete and highly combinatorial nature~\cite{hartmanis1982computers}, necessitating scalable and high-quality approximation algorithms for modern chip design.

\subsection{Motivations}

State-of-the-art (SOTA) hypergraph partitioners predominantly follow the multilevel paradigm~\cite{karypis1998hmetis,schlag2016k,liang2024medpart,sajadinia2025shypar,lee2025hyperg}. KaHyPar~\cite{schlag2016k}, for example, (1) \emph{coarsens} the hypergraph into successively smaller ones while attempting to preserve structure; (2) computes an \emph{initial partition} on the coarsest hypergraph; and (3) \emph{uncoarsens and refines} the solution at progressively finer levels. While highly effective, this workflow depends critically on coarsening quality. When coarsening distorts the original structure—particularly in hypergraphs with many high-degree edges—the refinement stage requires extensive local search to recover high-quality solutions, which can dominate runtime and limit scalability.

\begin{figure}[!t]
\centering
\vspace{-1mm}
\includegraphics[width=0.82\linewidth]{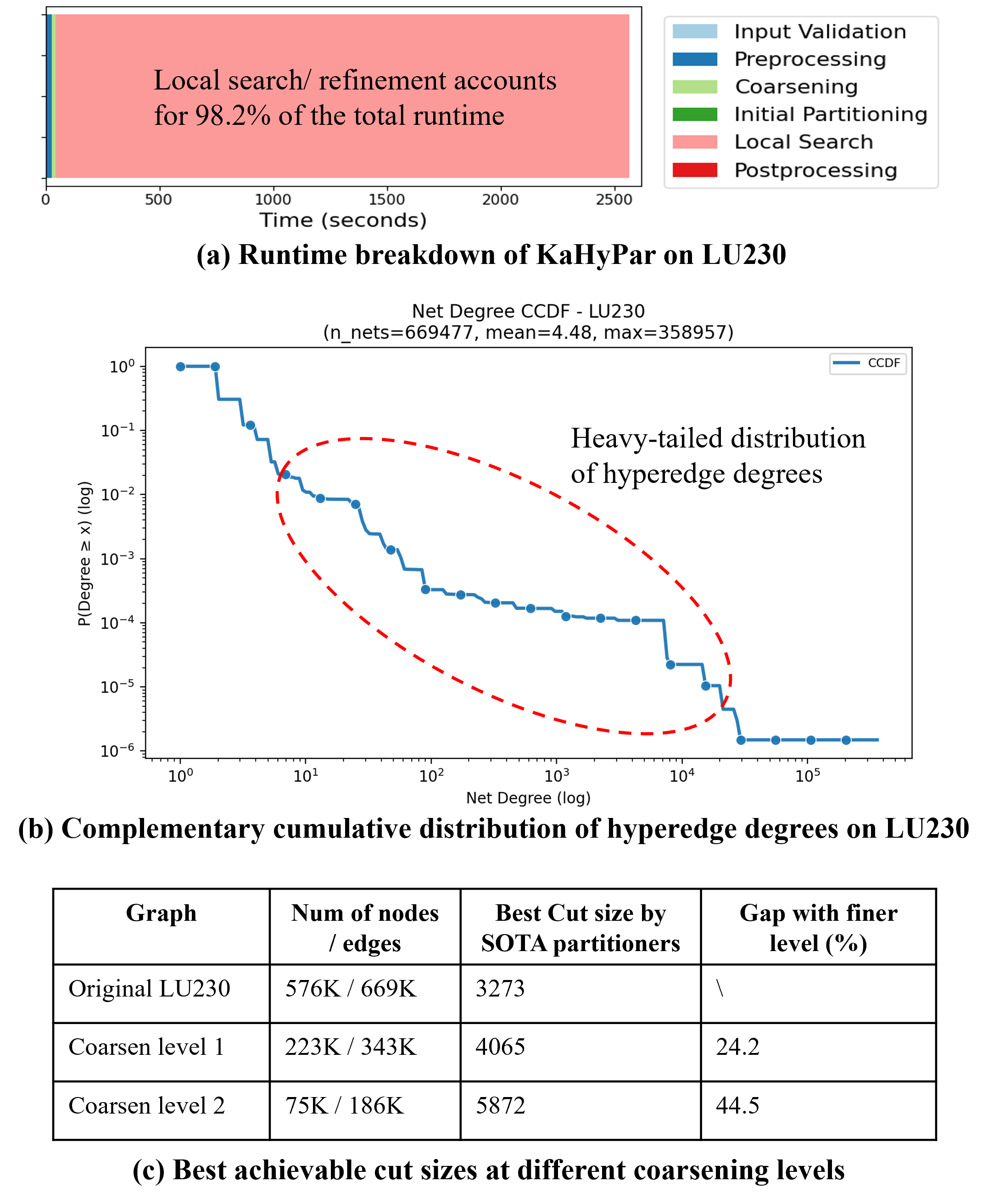}%
\vspace{-4.5mm}%
\caption{Analysis of LU230 and its partitioning results.}
\label{fig:example}
\vspace{-4.5mm}
\end{figure}

The LU230 testcase from the Titan23 suite~\cite{murray2013titan} exemplifies this issue. As shown in \Cref{fig:runtime}, KaHyPar requires nearly 2500 seconds on LU230, far slower than other similarly sized designs. The runtime breakdown in \Cref{fig:example}(a) shows that over 98\% of total time is spent in refinement. LU230 exhibits a heavy-tailed hyperedge-degree distribution (\Cref{fig:example}(b)), which makes structure-preserving coarsening particularly challenging. To quantify the resulting information loss, we compare the best achievable cut on the original hypergraph (estimated using several SOTA partitioners) with those obtained on the first two coarsened levels produced by KaHyPar. If coarsening preserved all essential structure, these cut values would be identical. Instead, \Cref{fig:example}(c) shows substantial discrepancies across levels, indicating that important structural information is lost during coarsening—thereby explaining the unusually large refinement cost on LU230.

Spectral techniques have proven highly effective in graph-based applications such as clustering, sparsification, and circuit modeling \cite{teng2016scalable, spielman2014sdd,  xueqian:iccad14, zhuo:dac18} by capturing global structural information through the spectrum of the Laplacian. While spectral graph theory enables nearly-linear algorithms for tasks like sparsification and coarsening \cite{spielman2011spectral, loukas2018spectrally}, analogous advances for hypergraphs are still in their early stages. Recent progress in nonlinear diffusion-based Laplacian formulations has provided rigorous theoretical foundations \cite{chan2018spectral, chan2020generalizing}, extending Cheeger’s inequality to hypergraphs, while practical implementations are still limited.

Recent methods~\cite{lee2025hyperg,liang2024medpart} have explored incorporating spectral information to mitigate coarsening distortion. But they remain bound to the multilevel pipeline where information loss is unavoidable. Moreover, coarsening is often largely oblivious to the true cost function and balance constraints, shifting these burdens to the refinement stage. Although such approaches often report improved cut quality, they rarely achieve better runtime. This raises a key question: \textbf{Can we step outside the multilevel framework and leverage spectral information more effectively to match the quality of spectral-guided multilevel partitioners while delivering substantially better scalability?}

A separate line of work performs single-level partitioning in a spectral embedding space. Classical spectral clustering partitions vertices using Laplacian eigenvectors but does not enforce balance, requiring substantial post-processing. 
SpecPart~\cite{bustany2023k, bustany2022specpart,bustany2023open} constructs spectral embeddings and refines high-quality initial solutions from multilevel methods using tree sweeps in the embedding space. However, it leverages spectral information indirectly and heuristically, making its cut quality heavily dependent on both the initial solutions and the tree construction heuristics.


\subsection{Contributions}

This work addresses the limitations of multilevel coarsening and the indirect use of spectral information by introducing a \emph{single-level} hypergraph partitioning paradigm based on \emph{end-to-end optimization} in a spectral embedding space. HySpecPro constructs the embedding using the first few Laplacian eigenvectors of the bipartite representation of the hypergraph, and then directly optimizes the partition via a projection-based continuous formulation.

Our main contributions are:
\begin{enumerate}[topsep=2pt, leftmargin=4.5mm]

\item \textbf{Single-level partitioning.}  
HySpecPro is, to our knowledge, the first single-level hypergraph partitioner to achieve cut quality competitive with SOTA multilevel methods.

\item \textbf{End-to-end continuous optimization.}  
We reformulate hypergraph partitioning as a continuous problem in the spectral embedding space and introduce an efficient projection-optimization procedure that directly targets the objective.

\item \textbf{GPU-accelerated implementation.}  
HySpecPro features a fully GPU-accelerated pipeline, employing CuPy~\cite{nishino2017cupy} for high-throughput eigenvector computation and Deep Graph Library (DGL)~\cite{wang2019deep} for batched evaluation of partition candidates.

\item \textbf{Strong empirical performance.}  
Experiments on the Titan23 and L\_HG suites demonstrate that HySpecPro delivers highly competitive cut sizes while scaling substantially better than leading multilevel methods, particularly on hypergraphs with many high-degree edges.

\item \textbf{Flexible, integrable framework.}
HySpecPro’s simple single-level, end-to-end formulation avoids the complex heuristics of prior partitioners, making it easy to adapt to objectives beyond min-cut and enabling much deeper integration with other optimization engines in VLSI design.

\item \textbf{Open-source release.}  
To support future research and integration with EDA workflows, we release HySpecPro as open source at \url{https://github.com/NVlabs/HySpecPro}.


\end{enumerate}

\section{PRELIMINARIES}

\begin{figure*}[!t]
\centering
\vspace{-2mm}
\includegraphics[width=0.7\linewidth]{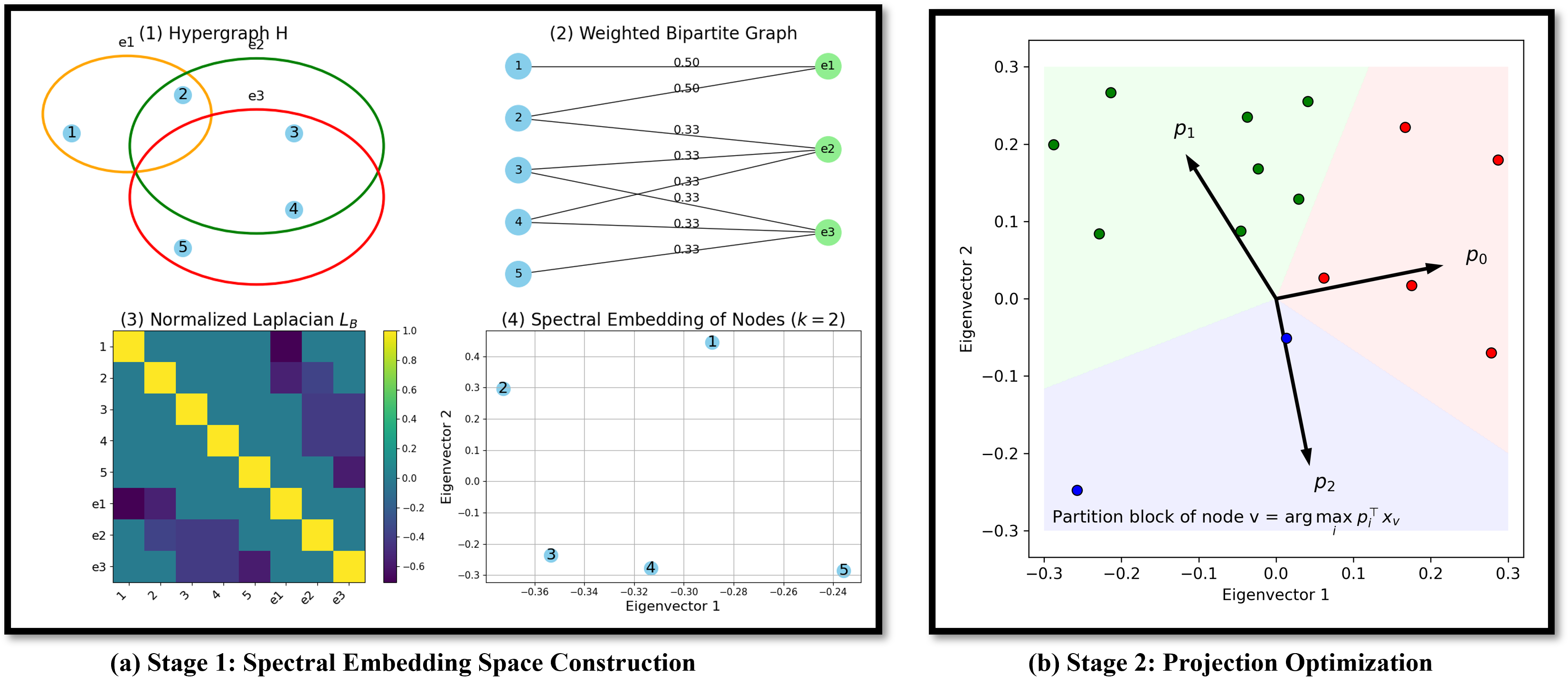}%
\vspace{-4mm}%
\caption{Illustration of the spectral projection optimization process in HySpecPro.}
\label{fig:overview}
\vspace{-4mm}
\end{figure*}

This section presents preliminaries necessary for the understanding of HySpecPro. We first offer the mathematical formulation for hypergraph partitioning, followed by a brief overview of the Covariance Matrix Adaptation Evolution Strategy used in HySpecPro.

\subsection{Hypergraph Partitioning Formulation}
Hypergraph partitioning seeks to divide the vertices of a hypergraph into
$K$ balanced blocks while minimizing the number of \emph{cut} hyperedges,
i.e., hyperedges whose pins lie in more than one block. Formally, let
$\mathcal{H} = (V, E)$ be a hypergraph with vertex weights $w(v) > 0$ and
hyperedge weights $\omega(e) > 0$. A partition of $V$ into
$K$ disjoint blocks $V_1,\dots,V_K$ induces, for each hyperedge $e$,
\[
\lambda_e \;=\; \bigl|\{\, i \mid e \cap V_i \neq \varnothing \,\}\bigr|,
\]
the number of blocks touched by $e$. The optimization objective is
\begin{equation}
\label{eq:objective}
  \min_{V_1,\dots,V_K} \; \mathrm{cut}(\mathcal{H})
    \;=\; \sum_{e \in E} \omega(e)\,\mathbf{1}\{\lambda_e > 1\},  
\end{equation}
where $\mathbf{1}\{\cdot\}$ is the indicator function.

Balance constraints with imbalance tolerance $\epsilon \ge 0$ require
\[
W_i = \sum_{v \in V_i} w(v), \qquad
W = \sum_{v \in V} w(v), \qquad
W_{\mathrm{avg}} = \frac{W}{K},
\]
\begin{equation}
W_{\mathrm{avg}} - \epsilon W \le\;  W_i \;\le\; W_{\mathrm{avg}} + \epsilon W, \qquad i = 1,\dots,K . 
\end{equation}

For simplicity but without loss of generality, this work focuses on the bipartitioning case
($K=2$) with unit vertex and hyperedge weights. Extensions to non-uniform
weights are left to future work.



\subsection{Covariance Matrix Adaptation Evolution Strategy (CMA-ES)}
CMA-ES~\cite{hansen2019pycma} is a stochastic, derivative-free 
optimization algorithm designed for difficult nonconvex and non-smooth search 
landscapes. At iteration $t$, the algorithm maintains a multivariate normal 
distribution $\mathcal{N}(m_t, \sigma_t^2 C_t)$ over the search space, where 
$m_t$ is the mean, $\sigma_t$ is a global step size, and $C_t$ is a full 
covariance matrix that captures the shape of the search distribution. A 
population of candidate solutions is sampled from this distribution, evaluated 
by the objective function, and ranked. The mean is then updated toward the 
weighted recombination of the best samples. The covariance matrix $C_t$ is 
adapted to increase variance along directions that consistently lead to better 
solutions and to reduce variance along unproductive directions, enabling 
automatic learning of the underlying problem geometry. The step size $\sigma_t$ 
is controlled separately through a cumulative path mechanism that encourages 
long-term progress while preventing premature convergence. CMA-ES is widely used 
because it is robust, requires no gradient information, adapts to ill-conditioned 
problems, and performs strongly on small-dimensional (e.g., 3--100) black-box optimization tasks.

\section{Methodology}

HySpecPro is a single-level hypergraph partitioning framework that performs end-to-end optimization in a spectral embedding space. It consists of two stages: (1) spectral embedding construction and (2) projection-based optimization, as shown in \Cref{fig:overview}. The following subsections detail these stages, analyze the runtime complexity, and describe two optional enhancements integrated into the framework.

\subsection{Spectral Embedding Space Construction}
\label{sec:spectral_embed}

HySpecPro constructs its embedding space by forming a weighted bipartite representation of the input hypergraph and computing the first few eigenvectors of its normalized Laplacian, as shown in \Cref{fig:overview}(a). These eigenvectors define the spectral embedding used for subsequent optimization.

\textbf{Weighted Bipartite Graph Construction.}
For a hypergraph $\mathcal{H} = (V, E)$, we build a weighted bipartite graph $\mathcal{B}$ with vertex nodes (one for each $v \in V$) and hyperedge nodes (one for each $e \in E$). An undirected edge connects $v$ and $e$ iff $v \in e$ in $\mathcal{H}$, with weight $\omega(e)/d(e)$, where $\omega(e)$ is the hyperedge weight and $d(e)$ its degree. This normalization prevents high-degree hyperedges from dominating the affinity structure.

\textbf{Eigenvector Computation.}
Let $A_{\mathcal{B}}$ denote the adjacency matrix of $\mathcal{B}$, and define its normalized Laplacian as
\[
L_{\mathcal{B}} = I - D_{\mathcal{B}}^{-1/2} A_{\mathcal{B}} D_{\mathcal{B}}^{-1/2},
\]
where $D_{\mathcal{B}}$ is the diagonal degree matrix. HySpecPro extracts the eigenvectors corresponding to the $S$ smallest nontrivial eigenvalues to obtain an $S$-dimensional spectral embedding of the vertex nodes. These eigenvectors capture global connectivity and form the continuous optimization space used by HySpecPro.

\subsection{Projection Optimization}
\label{sec:opt}

HySpecPro reformulates $K$-way hypergraph partitioning as the optimization of $K$ projection vectors in the spectral embedding space, as shown in \Cref{fig:overview}(b). These projections are then optimized using CMA--ES.

\textbf{Hypergraph Partitioning as Projection Optimization.}
Let \(X \in \mathbb{R}^{|V| \times S}\) be the spectral embedding matrix, where
row \(x_v^\top\) encodes the embedding of vertex \(v\). HySpecPro parameterizes
a \(K\)-way partition using a projection matrix
\[
P = [p_1,\dots,p_K] \in \mathbb{R}^{S \times K}.
\]

\paragraph{Assignment via projection.}
Each vertex is assigned to the block with the highest projection score:
\begin{equation}
\label{eq:argmax}
    y_v(P) = \arg\max_{k \in \{1,\dots,K\}} p_k^\top x_v.
\end{equation}

\paragraph{Hyperedge cut condition.}
A hyperedge \(e\) is cut if any two of its incident vertices receive different
assignments:
\begin{equation}
\label{eq:cut}
    z_e(P)=
    \begin{cases}
    1, & \exists\, v_i, v_j \in e \text{ s.t. } y_{v_i}(P) \neq y_{v_j}(P),\\[4pt]
    0, & \text{otherwise}.
    \end{cases}
\end{equation}
The partitioning objective (\Cref{eq:objective}) can therefore be written as
\begin{equation}
\label{eq:reformulate}
    \min_{P} \;\mathrm{cut}(\mathcal{H})
    \;=\; \sum_{e \in E} \omega(e)\, z_e(P),
\end{equation}
with standard balance constraints.

\paragraph{Compact continuous representation.}
Unlike the original discrete formulation over assignments
\((V_1,\dots,V_K)\), whose search space grows exponentially with \(|V|\), the
reformulation optimizes only the continuous matrix \(P\) with \(S \times K\)
parameters. This compact parameterization is central to HySpecPro’s scalable optimization.

\paragraph{Expressiveness of linear separators.}
The projection vectors induce hyperplanes that partition the spectral
embedding space (\Cref{fig:overview}(b)). Although the framework can be
extended to nonlinear separation boundaries, embeddings of netlist
hypergraphs in the Titan23 suite often exhibit a ray-like geometry
(\Cref{fig:embed}), making linear separation surprisingly effective.
Extending the framework to richer decision models is a promising direction for
future work.

\textbf{Continuous Optimization of Projection Vectors.}
HySpecPro optimizes the projection vectors using CMA--ES, a derivative-free
method well suited for this task. The objective in
\Cref{eq:argmax}–(\ref{eq:reformulate}) is highly nonconvex and piecewise
constant due to the winner-take-all assignment, making gradient-based
approaches unreliable. CMA--ES handles such landscapes robustly. Moreover,
real-world hypergraphs such as VLSI netlists exhibit strong community
structure, leading to spectral embeddings with small effective dimension
\(S\). As a result, the projection matrix \(P \in \mathbb{R}^{S \times K}\)
contains relatively few parameters, allowing CMA--ES to optimize it
efficiently.

In practice, CMA--ES is initialized at the origin of the embedding space, with
no explicit bounds on \(P\). Balance constraints are enforced through a penalty
term proportional to their degree of violation.

\subsection{GPU-Accelerated Implementation}

HySpecPro leverages GPU acceleration for both spectral embedding construction and projection optimization.

\textbf{GPU-Accelerated Spectral Embedding Construction.}
We use CuPy, a GPU-accelerated array library for Python, to form the normalized Laplacian and compute its first few eigenpairs. To reduce memory usage, the Laplacian and related matrices are stored in sparse format and processed using CuPy’s sparse linear-algebra operators.

\textbf{GPU-Accelerated Projection Optimization.}
The dominant cost in projection optimization is solution evaluation, as CMA--ES requires thousands of candidate evaluations per iteration. We adopt the GPU-accelerated batched evaluation strategy from MedPart~\cite{liang2024medpart}. Each candidate partition is encoded as node features on the bipartite graph~$\mathcal{B}$, represented as a heterogeneous graph in DGL. Cut sizes and balance constraints are computed through DGL message passing and PyTorch tensor operations, enabling efficient batched execution fully on the GPU.




\subsection{Runtime Complexity}
\label{sec:complexity}

HySpecPro computes the smallest $S$ eigenpairs of the normalized Laplacian
using CuPy’s GPU-accelerated Lanczos solver. For a sparse matrix with
$\mathrm{nnz}$ nonzeros, each Lanczos iteration is dominated by a sparse
matrix--vector multiplication costing $\mathcal{O}(\mathrm{nnz})$. As defined in
\Cref{sec:spectral_embed}, the normalized Laplacian $L_{\mathcal{B}}$ has
$\mathrm{nnz} = \mathcal{O}(\mathrm{TotalDeg})$, where $\mathrm{TotalDeg}$ is
the total hyperedge degree of $\mathcal{H}$. Computing $S$ eigenpairs therefore
requires
\[
\mathcal{O}\!\bigl(\mathrm{TotalDeg}\cdot S \cdot I\bigr),
\]
where $I$ is the number of Lanczos iterations. Although GPU acceleration does
not change the asymptotic cost, it substantially reduces wall-clock time via
high-throughput sparse linear algebra.

In the projection-optimization stage, the dominant cost is evaluating candidate
projection matrices. Each evaluation performs a DGL message-passing pass on the
bipartite graph, with complexity $\mathcal{O}(\mathrm{TotalDeg})$. With a
CMA--ES population size $B$ and $T$ generations, the total cost is
\[
\mathcal{O}\!\left(\mathrm{TotalDeg}\cdot B \cdot T\right).
\]
DGL’s GPU backend parallelizes edgewise operations, improving runtime in
practice but preserving asymptotic behavior.

Treating $S$, $I$, $B$, and $T$ as constants, the overall runtime of HySpecPro
scales linearly with $\mathrm{TotalDeg}$.




\subsection{Optional Enhancements}
\label{sec:enhancement}

Two optional enhancements—hybrid spectral embeddings and differentiable refinement—can be enabled in HySpecPro when partitioning quality is prioritized over runtime.

\textbf{Hybrid Spectral Embeddings.}
In addition to the bipartite graph, HySpecPro can construct a \emph{sampled clique-expansion graph} to capture higher-order relationships within hyperedges. A full clique expansion replaces each hyperedge of size $d(e)$ with all $\binom{d(e)}{2}$ pairwise edges, which is impractical for large hyperedges. To reduce memory and runtime, we instead sample a subset of these pairwise edges for high-degree hyperedges, preserving structural information while keeping the graph sparse.
The spectral-embedding procedure in \Cref{sec:spectral_embed} is then applied to this sampled clique-expansion graph to obtain its first few eigenvectors. HySpecPro evaluates both the bipartite and clique-based embeddings by running CMA--ES on each and selects the better resulting partition.

\textbf{Differentiable Refinement.}
While the smallest nontrivial eigenvectors capture global structure, some local detail may be missed. As an optional post-processing step, HySpecPro can switch back to the original solution space and apply the differentiable refinement method from MedPart to further improve the partition.

\begin{table}[t]
\vspace{-4mm}
\caption{Statistics of the Titan23 benchmark suite.}
\vspace{-4mm}
\centering
\scriptsize
\begin{tabular}{lrrrr}
\hline
\textbf{Benchmark} & \multicolumn{1}{c}{$|V|$} & \multicolumn{1}{c}{$|E|$} & \multicolumn{1}{c}{TotalDeg} & \multicolumn{1}{c}{Sum\_deg\_ge\_100} \\
\hline
sparcT1 core      & 91{,}976   & 92{,}827    & 503{,}922   & 138{,}351 \\
neuron            & 92{,}290   & 125{,}305   & 453{,}181   & 127{,}232 \\
stereo vision     & 94{,}050   & 127{,}085   & 414{,}259   & 91{,}078  \\
des90             & 111{,}221  & 139{,}557   & 771{,}009   & 96{,}847  \\
SLAM spheric      & 113{,}115  & 142{,}408   & 637{,}880   & 37{,}343  \\
cholesky mc       & 113{,}250  & 144{,}948   & 622{,}678   & 201{,}216 \\
segmemtation      & 138{,}295  & 179{,}051   & 801{,}427   & 43{,}795  \\
bitonic mesh      & 192{,}064  & 235{,}328   & 1{,}364{,}768 & 186{,}557 \\
dart              & 202{,}354  & 223{,}301   & 1{,}133{,}096 & 254{,}433 \\
openCV            & 217{,}453  & 284{,}108   & 1{,}273{,}473 & 305{,}429 \\
stap qrd          & 240{,}240  & 290{,}123   & 1{,}260{,}458 & 391{,}561 \\
minres            & 261{,}359  & 320{,}540   & 1{,}309{,}022 & 250{,}992 \\
cholesky bdti     & 266{,}422  & 342{,}688   & 1{,}410{,}151 & 446{,}869 \\
denoise           & 275{,}638  & 356{,}848   & 1{,}610{,}717 & 118{,}249 \\
sparcT2 core      & 300{,}109  & 302{,}663   & 1{,}615{,}849 & 501{,}749 \\
gsm switch        & 493{,}260  & 507{,}821   & 2{,}712{,}782 & 1{,}213{,}869 \\
mes noc           & 547{,}544  & 577{,}664   & 3{,}187{,}553 & 936{,}686 \\
LU230             & 574{,}372  & 669{,}477   & 2{,}996{,}579 & 1{,}123{,}810 \\
LU Network        & 635{,}456  & 726{,}999   & 3{,}331{,}887 & 801{,}400 \\
sparcT1 chip2     & 820{,}886  & 821{,}274   & 4{,}031{,}921 & 1{,}317{,}794 \\
directrf          & 931{,}275  & 1{,}374{,}742 & 4{,}812{,}506 & 1{,}007{,}163 \\
bitcoin miner     & 1{,}089{,}284 & 1{,}448{,}151 & 4{,}557{,}197 & 907{,}382 \\
\hline
\end{tabular}
\label{tab:titan23_sta}
\vspace{-6mm}
\end{table}

\section{Experimental Results and Analysis}

\begin{table}[t]
\vspace{-3mm}
\caption{Bi-partitioning cut sizes obtained by different approaches on the Titan23 benchmark suite. The hMETIS, SpecPart, MedPart, SHyPar and KaHyPar results are copied from~\cite{sajadinia2025shypar}. The average improvement is normalized against hMETIS results. }
\label{tab:titan-cut}
\vspace{-3mm}
\centering
\scriptsize
\setlength{\tabcolsep}{2pt}
\renewcommand{\arraystretch}{1.05}
\begin{tabular}{l | r r r r r r r r| r r r}
\toprule
\multirow{2}{*}{\textbf{Benchmark}} &
\multicolumn{10}{c}{\textbf{Cut Size at $\epsilon=2\%$}} \\
\cmidrule(lr){2-12}
& \rotatebox{90}{\textbf{hMETIS}}
& \rotatebox{90}{\textbf{SpecPart}}
& \rotatebox{90}{\textbf{MedPart}}
& \rotatebox{90}{\textbf{SHyPar}}
& \rotatebox{90}{\textbf{KaHyPar}}
& \rotatebox{90}{\textbf{TritonPart}}
& \rotatebox{90}{\textbf{mtK-min}}
& \rotatebox{90}{\textbf{mtK-mean}}
& \rotatebox{90}{\textbf{HySP}}
& \rotatebox{90}{\textbf{HySP-Hyb}}
& \rotatebox{90}{\textbf{HySP-RF}} \\
\midrule

sparcT1 core & 1066 & 1012 & 1067 & \textcolor{red}{974} & \textcolor{red}{974} & 1137 & 1099 & 1288 & 1091 & 1078 & 1048 \\
neuron & 260 & 252 & 262 & \textcolor{red}{243} & 244 & 245 & 260 & 380 & 248 & 249 & 247 \\
stereo vision & 180 & 180 & 176 & \textcolor{red}{169} & \textcolor{red}{169} & 182 & 174 & 254 & 177 & 171 & 178 \\
des90 & 402 & 402 & \textcolor{red}{372} & 379 & 380 & 382 & 389 & 477 & 380 & 377 & 375 \\
SLAM spheric & \textcolor{red}{1061} & \textcolor{red}{1061} & \textcolor{red}{1061} & \textcolor{red}{1061} & \textcolor{red}{1061} & \textcolor{red}{1061} & \textcolor{red}{1061} & 1083 & \textcolor{red}{1061} & \textcolor{red}{1061} & \textcolor{red}{1061} \\
cholesky mc & 285 & 285 & 283 & 283 & 283 & \textcolor{red}{282} & 382 & 756 & 283 & \textcolor{red}{282} & \textcolor{red}{282} \\
segmentation & 136 & 126 & 114 & \textcolor{red}{107} & \textcolor{red}{107} & \textcolor{red}{107} & 144 & 256 & 128 & 120 & 118 \\
bitonic mesh & 614 & \textcolor{red}{585} & 594 & 586 & 593 & 588 & 590 & 671 & 589 & 594 & 587 \\
dart & 844 & 807 & 805 & \textcolor{red}{784} & 924 & 846 & 817 & 1113 & 809 & 837 & 797 \\
openCV & 511 & 510 & 635 & 499 & 560 & \textcolor{red}{429} & 591 & 798 & 575 & 588 & 576 \\
stap qrd & 399 & 399 & 386 & \textcolor{red}{371} & \textcolor{red}{371} & 382 & 380 & 509 & 379 & 384 & 379 \\
minres & 215 & 215 & 215 & \textcolor{red}{207} & \textcolor{red}{207} & 208 & 252 & 351 & 210 & 210 & 213 \\
cholesky bdti & 1157 & \textcolor{red}{1156} & 1161 & \textcolor{red}{1156} & \textcolor{red}{1156} & 1197 & 1209 & 1792 & 1157 & 1157 & 1157 \\
denoise & 722 & \textcolor{red}{416} & 516 & \textcolor{red}{416} & \textcolor{red}{416} & 479 & 456 & 533 & 439 & 428 & 546 \\
sparcT2 core & 1273 & 1244 & 1319 & \textcolor{red}{1183} & 1186 & 1310 & 1196 & 1301 & 1377 & 1373 & 1281 \\
gsm switch & 5974 & 1827 & 1714 & 1621 & 1759 & \textcolor{red}{1496} & 2036 & 2789 & 1590 & 1663 & 1705 \\
mes noc & 699 & \textcolor{red}{634} & 699 & 651 & 649 & 886 & 673 & 935 & 646 & 646 & 646 \\
LU230 & 4070 & \textcolor{red}{3273} & 3452 & 3602 & 4012 & 3426 & 3614 & 4390 & 3399 & 3410 & 3320 \\
LU Network & 550 & 525 & 550 & \textcolor{red}{524} & \textcolor{red}{524} & 528 & 610 & 737 & 536 & 533 & 542 \\
sparcT1 chip2 & 1524 & 899 & 1129 & \textcolor{red}{873} & 874 & 1043 & 911 & 1196 & 903 & 910 & 907 \\
directrf & 646 & 574 & 646 & 632 & 646 & 576 & 662 & 731 & 506 & \textcolor{red}{505} & 508 \\
bitcoin miner & 1570 & \textcolor{red}{1297} & 1562 & 1514 & 1576 & 1489 & 1523 & 1582 & 1651 & 1582 & 1618 \\
\midrule
\textbf{Impro. \% $\uparrow$} 
& 0 & 10.99 & 6.82 & \textcolor{red}{12.16} & 9.95 & 8.90 & 4.16 & -28.00 & 9.42 & 9.80 & 9.70 \\
\bottomrule
\end{tabular}

\vspace{0.4em}

\begin{tabular}{l | r r r r r r r r | r r r}
\toprule
\multicolumn{12}{c}{\textbf{Cut Size at $\epsilon=20\%$}} \\
\midrule
sparcT1 core & 1290 & 903 & 624 & 631 & 873 & 1016 & 1064 & 1210 & 682 & 643 & \textcolor{red}{598} \\
neuron & 270 & 206 & 270 & 244 & 244 & 245 & 251 & 360 & 216 & \textcolor{red}{205} & 209 \\
stereo vision & 143 & \textcolor{red}{91} & 93 & \textcolor{red}{91} & \textcolor{red}{91} & \textcolor{red}{91} & \textcolor{red}{91} & 103 & 97 & 96 & 95 \\
des90 & 441 & 358 & 349 & 345 & 380 & 353 & 382 & 481 & 345 & 329 & \textcolor{red}{327} \\
SLAM spheric & \textcolor{red}{1061} & \textcolor{red}{1061} & \textcolor{red}{1061} & \textcolor{red}{1061} & \textcolor{red}{1061} & \textcolor{red}{1061} & \textcolor{red}{1061} & 1076 & \textcolor{red}{1061} & \textcolor{red}{1061} & \textcolor{red}{1061} \\
cholesky mc & 667 & 345 & \textcolor{red}{281} & 479 & 591 & \textcolor{red}{281} & \textcolor{red}{281} & 671 & \textcolor{red}{281} & \textcolor{red}{281} & \textcolor{red}{281} \\
segmentation & 141 & \textcolor{red}{78} & \textcolor{red}{78} & \textcolor{red}{78} & \textcolor{red}{78} & 80 & 86 & 148 & 82 & 80 & \textcolor{red}{78} \\
bitonic mesh & 590 & \textcolor{red}{483} & 493 & 506 & 493 & 506 & 588 & 675 & 489 & 506 & \textcolor{red}{483} \\
dart & 603 & 540 & 549 & \textcolor{red}{539} & 549 & 583 & 797 & 1127 & 667 & 666 & 547 \\
openCV & 554 & 518 & 554 & 473 & 554 & 451 & 625 & 778 & 534 & \textcolor{red}{450} & 503 \\
stap qrd & 295 & 295 & 287 & \textcolor{red}{275} & 287 & 281 & 364 & 490 & 276 & 276 & 276 \\
minres & 189 & 189 & \textcolor{red}{181} & 191 & \textcolor{red}{181} & 211 & 220 & 317 & 182 & 183 & 185 \\
cholesky bdti & 1024 & 947 & 1024 & \textcolor{red}{848} & 1024 & 861 & 1245 & 1680 & 858 & 861 & 850 \\
denoise & 478 & 224 & 224 & \textcolor{red}{220} & 224 & 231 & 340 & 436 & 284 & 298 & 251 \\
sparcT2 core & 1972 & 1245 & 1081 & \textcolor{red}{918} & 1081 & 1171 & 1192 & 1298 & 999 & 994 & 974 \\
gsm switch & 5352 & \textcolor{red}{1407} & 1503 & \textcolor{red}{1407} & 1503 & 1695 & 1641 & 2201 & 1434 & 1439 & 1430 \\
mes noc & 633 & \textcolor{red}{617} & 633 & \textcolor{red}{617} & 633 & 623 & 634 & 860 & 631 & 629 & 629 \\
LU230 & 3276 & \textcolor{red}{2677} & 2720 & 2923 & 2720 & 3529 & 3586 & 4238 & 3017 & 2876 & 2790 \\
LU Network & 528 & \textcolor{red}{524} & 528 & \textcolor{red}{524} & 528 & 529 & 526 & 532 & 542 & 541 & 537 \\
sparcT1 chip2 & 1029 & 783 & 877 & 757 & 877 & 939 & 847 & 1028 & \textcolor{red}{735} & 738 & 791 \\
directrf & 379 & \textcolor{red}{295} & 317 & \textcolor{red}{295} & 317 & \textcolor{red}{295} & 385 & 531 & 318 & 316 & 319 \\
bitcoin miner & 1255 & \textcolor{red}{1225} & 1255 & 1282 & 1255 & 1278 & 1232 & 1514 & 1231 & 1233 & 1231 \\
\midrule
\textbf{Impro. \% $\uparrow$} 
& 0 & 21.73 & 20.94 & 22.53 & 14.14 & 18.93 & 9.62 & -17.26 & 21.48 & 22.69 & \textcolor{red}{23.93} \\
\bottomrule
\end{tabular}
\end{table}


HySpecPro is implemented in Python. All experiments are conducted on a server with AMD EPYC 7742 CPUs and an NVIDIA A100 GPU (80 GB). We use $S=32$ eigenvectors, $I=50$ Lanczos iterations, and run CMA-ES with a population size $B=3000$ or the maximum size allowed by GPU memory. The number of evolutionary generations is set to $T=50$.

We evaluate three configurations of HySpecPro:
\begin{itemize}[topsep=2pt, leftmargin=4.5mm]
    \item \textbf{HySP}: the baseline configuration of HySpecPro;
    \item \textbf{HySP-Hyb}: with Hybrid Spectral Embeddings enabled;
    \item \textbf{HySP-RF}: with Differentiable Refinement enabled.
\end{itemize}
Although HySP-Hyb and HySP-RF are theoretically guaranteed to achieve cut sizes no worse than HySP, their results may occasionally be slightly worse in practice due to the nondeterministic behavior of CMA-ES.

We compare them against the following SOTA partitioners:
\begin{itemize}[topsep=2pt, leftmargin=4.5mm]
    \item \textbf{hMETIS}~\cite{karypis1998hmetis}: a widely used sequential multilevel partitioner;
    \item \textbf{SpecPart}~\cite{bustany2022specpart}: a refinement method that improves hMETIS using spectral embeddings;
    \item \textbf{MedPart}~\cite{liang2024medpart}: a GPU-accelerated multilevel evolutionary differentiable partitioner;
    \item \textbf{KaHyPar}~\cite{schlag2016k}: a high-quality sequential multilevel partitioner with advanced refinement;
    \item \textbf{SHyPar}~\cite{sajadinia2025shypar}: an extension of KaHyPar that incorporates spectral information during coarsening;
    \item \textbf{TritonPart}~\cite{bustany2023open}: a leading multi-threaded partitioner; we use its official implementation~\cite{triton} with default settings;
    \item \textbf{mtK-min}~\cite{gottesburen2024scalable}: a multi-threaded KaHyPar configuration (also named \textbf{mtKaHyPar}, 24 threads); we report the best cut and the total runtime over 100 runs;
    \item \textbf{mtK-mean}~\cite{gottesburen2024scalable}: the same configuration, reporting average cut size and runtime over 100 runs.
\end{itemize}
To assess performance and scalability, we apply HySpecPro to the Titan23 benchmarks~\cite{murray2013titan} and eight large, high-complexity hypergraphs from the L\_HG suite~\cite{gottesburen2024scalable}, characterized by high \textit{TotalDeg} and many large hyperedges. Dataset statistics are provided in \Cref{tab:titan23_sta} (where \textit{Sum\_deg\_ge\_100} denotes the total degree contributed by hyperedges of degree $>100$) and \Cref{tab:LHG}. The following subsections present detailed results and analysis.

\begin{table*}[!bth]
\centering
\vspace{-3mm}
\caption{Statistics of L\_HG benchmark suite and bi-partitioning cut sizes obtained by different approaches.}
\label{tab:LHG}
\vspace{-3mm}
\scriptsize
\setlength{\tabcolsep}{4pt}
\renewcommand{\arraystretch}{1.15}
\begin{tabular}{lrrrrrrrrrrrrrr}
\toprule
\multirow{2}{*}{\textbf{Benchmark}} &
\multicolumn{3}{c}{\textbf{Statistics}} &
\multicolumn{5}{c}{\textbf{$\epsilon = 2\%$}} &
\multicolumn{5}{c}{\textbf{$\epsilon = 10\%$}} \\
\cmidrule(lr){2-4} \cmidrule(lr){5-9} \cmidrule(lr){10-14}
 & \(|V|\) & \(|E|\) & TotalDeg &
\textbf{KaHyPar} &
\textbf{TritonPart} &
\textbf{mtK-min} &
\textbf{mtK-mean} &
\textbf{HySP} &
\textbf{KaHyPar} &
\textbf{TritonPart} &
\textbf{mtK-min} &
\textbf{mtK-mean} &
\textbf{HySP} \\
\midrule

StocF-1465    & 1,465,137 & 1,465,137 & 21,005,389 & \textcolor{red}{4595} & 4812 & 4668 & 4879 & 4644 & \textcolor{red}{4595} & 4742 & 4667 & 4819 & 4597 \\
HV15R         & 2,017,169 & 2,017,169 & 283,073,458 & \textcolor{red}{60057} & 60747 & \textcolor{red}{60057}  & 60448 & 60060 & \textcolor{red}{60057} & 60721 & \textcolor{red}{60057} & 61172 & 60060 \\
Geo\_1438     & 1,437,960 & 1,437,960 & 63,156,690  & \textcolor{red}{26358} & 29505 & 26826 & 28498 & 27777 & \textcolor{red}{26088} & 30432 & 26700 & 28071 & 27285 \\
dgreen        & 1,200,611 & 1,200,611 & 38,259,877  & 9243 & \textcolor{red}{8414} & 8886 & 9402 & 8909 & 7684 & 7197 & 7530 & 7994 & \textcolor{red}{6894} \\
Ga41As41H72   & 268,096   & 268,096   & 18,488,476 & 36984 & 38048 & 37473 & 43205 & \textcolor{red}{36293} & 37302 & 41124 & 37111 & 42893 & \textcolor{red}{36172} \\
Bump\_2911    & 2,911,419 & 2,911,419 & 26,515,867  & \textcolor{red}{58815} & 61488 & 59028 & 59999 & 59081 & TIMEOUT & 60441 & \textcolor{red}{58866} & 59393 & 58980 \\
CurlCurl\_4   & 2,380,515 & 2,380,515 & 127,729,899 & TIMEOUT & 28186 & 25818 & 27507 & \textcolor{red}{25093} & TIMEOUT & 24087 & 24473 & 26028 & \textcolor{red}{22057} \\
circuit5M     & 5,558,326 & 5,558,326 & 59,524,291  & TIMEOUT & TIMEOUT & \textcolor{red}{9088} & 32202 & 9236 & TIMEOUT & 6523 & \textcolor{red}{6405} & 8191 & 6522 \\
\midrule

\multicolumn{4}{l}{\textbf{Cut size increase over HySP (\%)} $\downarrow$} &
NA & NA & 0.16 & 36.59 & 0 &
NA & 5.70 & 2.52 & 11.05 & 0 \\

\bottomrule
\end{tabular}
\end{table*}

\begin{figure}[!bth]
\centering
\begin{minipage}[t]{0.48\linewidth}
    \centering
    \includegraphics[width=\linewidth]{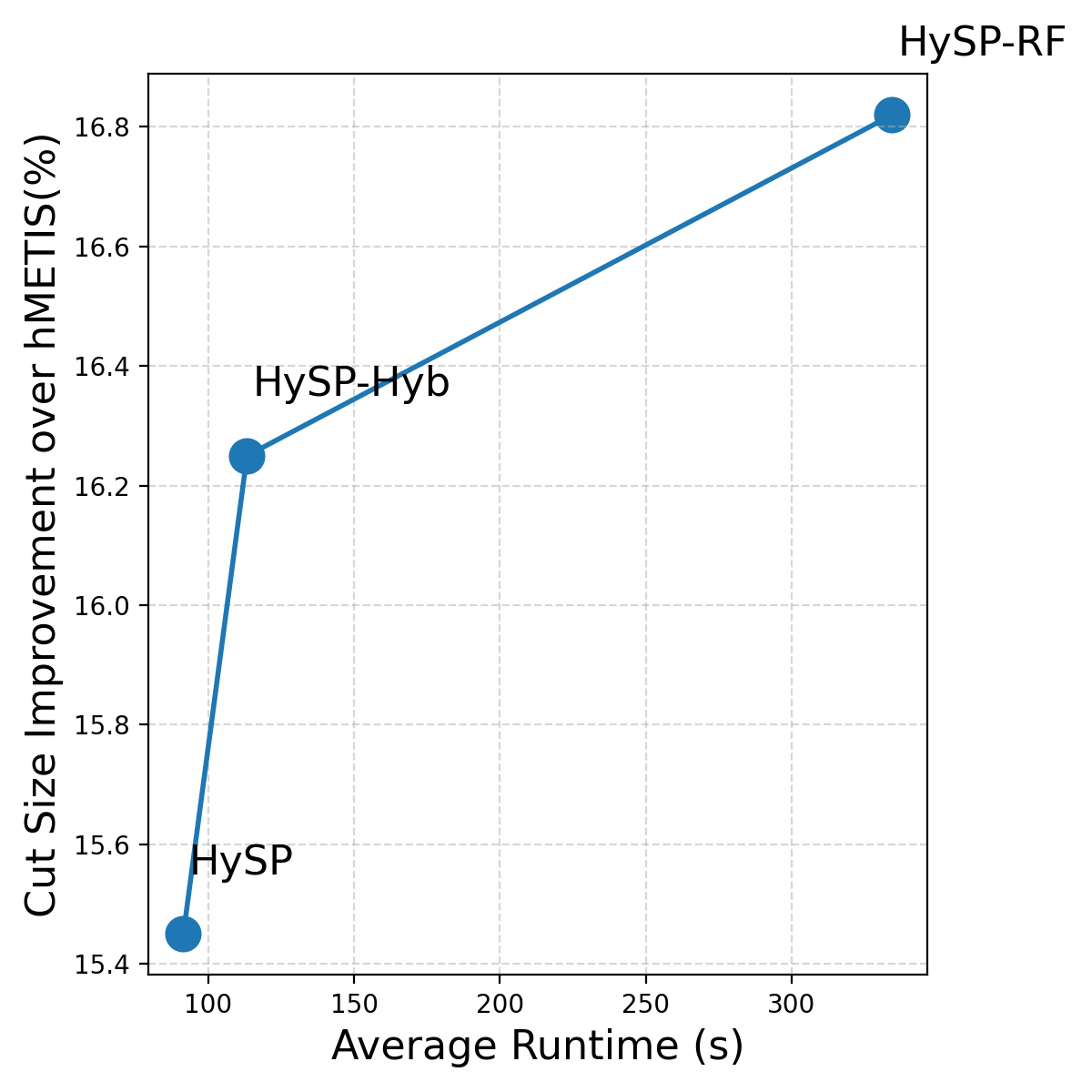}
    \vspace{-7mm}
    \caption{Runtime versus average cut-size improvement over hMETIS, aggregated across all Titan23 benchmarks for $\epsilon = 2\%$ and $\epsilon = 20\%$.}
    \label{fig:tradeoff}
\end{minipage}
\hfill
\begin{minipage}[t]{0.48\linewidth}
    \centering
    \includegraphics[width=\linewidth]{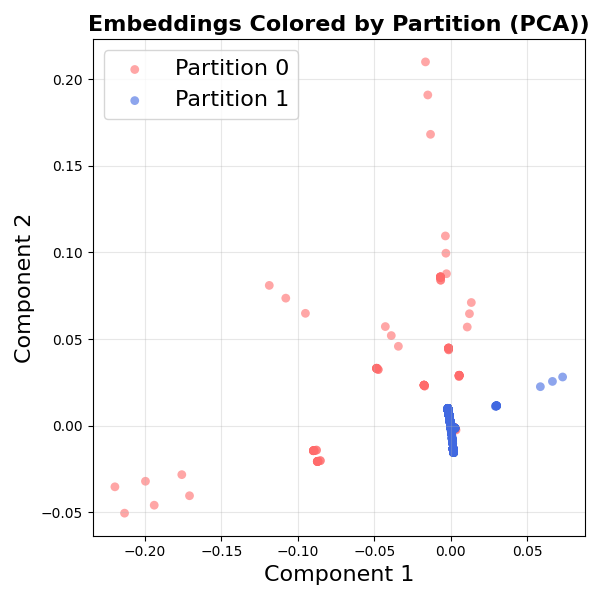}
    \vspace{-7mm}
    \caption{Visualization of the spectral embeddings (projected to 2D using PCA) and the resulting partitions produced by HySpecPro on sparcT1 core.}
    \label{fig:embed}
\end{minipage}
\vspace{-3mm}
\end{figure}

\begin{figure*}[!bth]
\centering
\vspace{-2.5mm}
\includegraphics[width=0.97\linewidth]{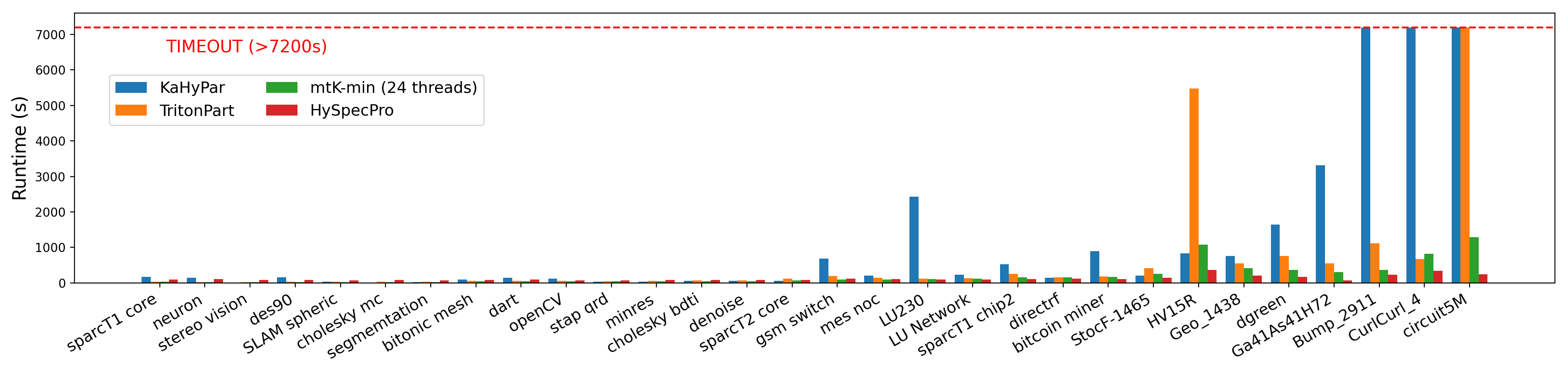}%
\vspace{-5mm}%
\caption{Runtime comparison of the four partitioners on the Titan23 and L\_HG benchmarks, aggregated over different $\epsilon$ settings.}
\vspace{-2mm}
\label{fig:runtime}
\end{figure*}

\begin{figure*}[!bth]
\centering
\vspace{-2mm}
\begin{minipage}[t]{0.48\linewidth}
    \centering
    \includegraphics[width=\linewidth]{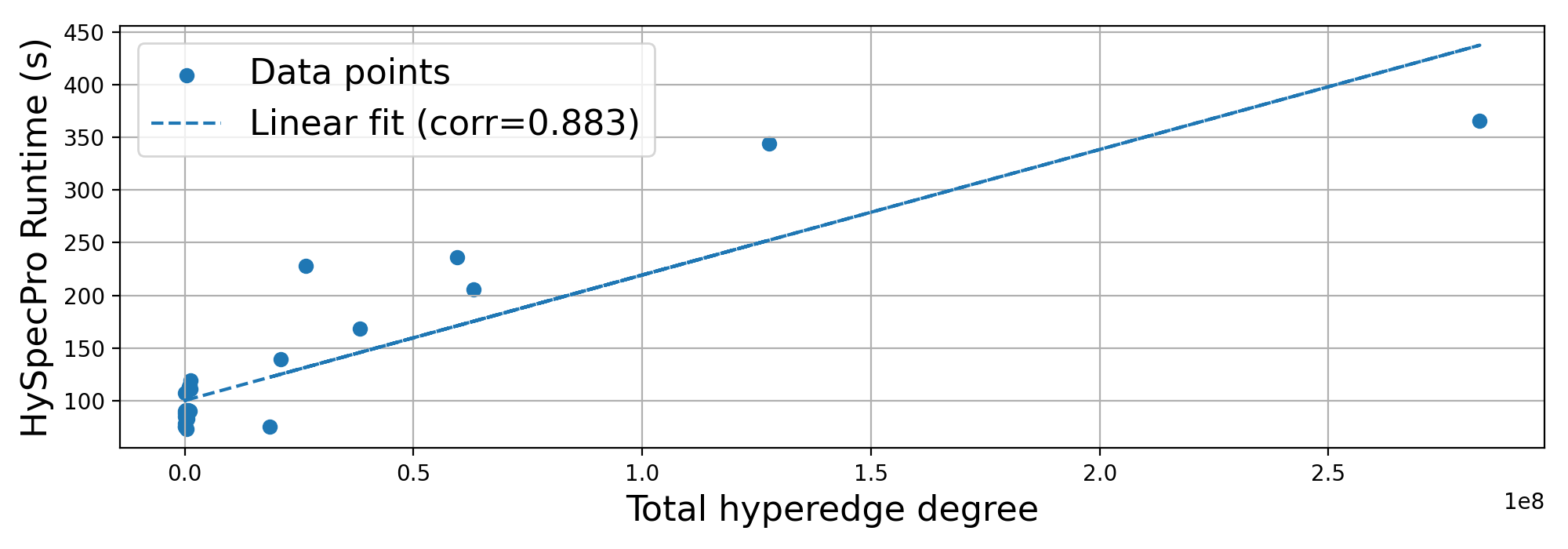}
    \vspace{-7mm}
    \caption{Runtime scalability of HySpecPro.}
    \label{fig:scalability}
\end{minipage}
\hfill
\begin{minipage}[t]{0.48\linewidth}
    \centering
    \includegraphics[width=\linewidth]{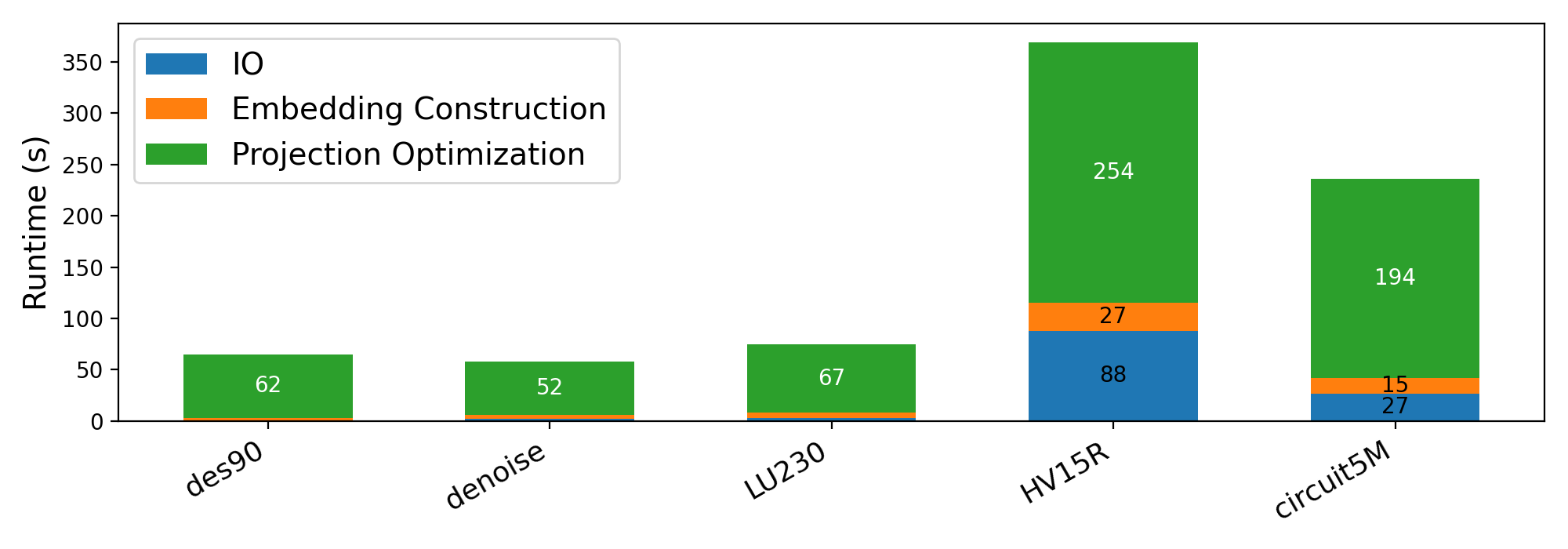}
    \vspace{-7mm}
    \caption{Runtime breakdown of HySpecPro.}
    \label{fig:breakdown}
\end{minipage}
\vspace{-3mm}
\end{figure*}

\subsection{Results on Titan23 Benchmarks}

\Cref{tab:titan-cut} summarizes the cut sizes on the Titan benchmarks. Across the suite, all three HySpecPro variants achieve competitive cut sizes, outperforming hMETIS by $9.42\%$–$9.80\%$ at $\epsilon = 2\%$ and by $21.48\%$–$23.93\%$ at $\epsilon = 20\%$ on average. These gains are highly comparable with previously best-reported improvements of $12.16\%$ and $22.53\%$, respectively. HySpecPro also attains new best cut sizes on several benchmarks, as indicated in \Cref{tab:titan-cut}.

In terms of runtime, SpecPart, which refines hMETIS partitions, incurs an approximately $5\times$ runtime overhead relative to hMETIS~\cite{bustany2023k}. MedPart requires up to two hours on the largest Titan benchmarks~\cite{liang2024medpart}, while SHyPar~\cite{sajadinia2025shypar} reports runtimes similar to KaHyPar. Detailed runtime comparisons among KaHyPar, TritonPart, mtKaHyPar (24 threads), and HySpecPro are presented in \Cref{fig:runtime}. Although HySpecPro is slower than KaHyPar and TritonPart on small hypergraphs, it is consistently faster than them when \textit{TotalDeg} exceeds $2{,}000{,}000$. The speedups of HySpecPro relative to KaHyPar are especially pronounced for instances with large \textit{Sum\_deg\_ge\_100}, such as \textit{gsm switch}, \textit{LU230}, and \textit{bitcoin miner}, where HySpecPro achieves $5.6\times$, $27\times$, and $8.09\times$ accelerations, respectively.
For mtKaHyPar, we report the mtK-min runtime (the total across 100 runs) rather than mtK-mean, since the latter produces substantially worse cuts than HySpecPro. Relative to mtK-min, HySpecPro is generally slower on this suite but becomes faster on benchmarks with large \textit{Sum\_deg\_ge\_100}, while consistently delivering superior cut quality.

\Cref{fig:tradeoff} illustrates the runtime–cut-size tradeoff among HySpecPro variants. The baseline HySpecPro is the fastest; HySpecPro-Hyb provides an additional $0.8\%$ cut reduction over hMETIS at a $24\%$ runtime increase, and HySpecPro-RF delivers about $1.4\%$ further improvement at a $3.6\times$ runtime cost.

\subsection{Results on L\_HG Benchmarks}
\Cref{tab:LHG} and \Cref{fig:runtime} report cut sizes and runtimes of KaHyPar, TritonPart, mtKaHyPar, and HySpecPro on the L\_HG benchmarks. In this regime, KaHyPar and TritonPart are significantly slower. For example, KaHyPar reaches the 7200\,s timeout on three benchmarks that HySpecPro completes in under 350\,s, while TritonPart is on average $7.3\times$ slower than HySpecPro.
In terms of cut quality, KaHyPar and HySpecPro achieve comparable results. HySpecPro outperforms TritonPart in nearly all cases, reducing cut size by an average of $5.7\%$. Compared to mtKaHyPar, HySpecPro is consistently faster across all benchmarks while also achieving smaller average cuts.




\subsection{Comparison of Spectral Embeddings}
By replacing the bipartite weights in \Cref{sec:spectral_embed} from $\omega(e)/d(e)$ to $\omega(e)$, we observe an average $5\%$ reduction in cut-size improvement on the Titan23 benchmarks at $\epsilon=20\%$, underscoring the importance of using normalized bipartite weights. 
In HySpecPro-Hyb, we further examine which embedding—induced by weighted bipartite or by sampled clique-expansion—produces the better partition. We find that the weighted bipartite embedding yields the superior result in roughly $95\%$ of the cases, providing additional evidence for the effectiveness of the spectral embedding design in \Cref{sec:spectral_embed}.

\subsection{Comparison with Alternative Optimization Techniques in the Embedding Space}
We also implement a differentiable alternative for optimizing the projection vectors (\Cref{sec:opt}) by replacing the winner-take-all assignment in \Cref{eq:argmax} with a softmax-based relaxation and substituting the objective in \Cref{eq:reformulate} with the differentiable proxy from MedPart. Relative to the CMA--ES approach, this method yields an average $4\%$ degradation in cut-size improvement on the Titan23 benchmarks at $\epsilon = 20\%$.

CuGraph’s \texttt{spectralBalancedCutClustering} function~\cite{fender2022rapids} provides a spectral balanced-cut clustering method for graphs but does not support hypergraphs. To enable comparison, we apply it to a sampled clique-expansion of each hypergraph. The resulting partitions are frequently imbalanced, and even after applying the differentiable refinement described in \Cref{sec:enhancement}, their quality remains inferior to that of HySpecPro.

\subsection{Runtime Analysis}

\Cref{fig:scalability} shows that HySpecPro’s runtime grows linearly with $\mathrm{TotalDeg}$, consistent with the theoretical analysis in \Cref{sec:complexity}. \Cref{fig:breakdown} provides a runtime breakdown on 5 representative benchmarks: projection optimization dominates the total runtime, whereas GPU-accelerated embedding construction finishes within 27\,s even for hypergraphs with $\mathrm{TotalDeg}$ up to 283 million. I/O overhead is also noticeable due to our current naive Python implementation, and could be substantially reduced with a more optimized pipeline.




\section{Conclusions}

In this work, we introduced HySpecPro, the first single-level hypergraph partitioner to achieve cut sizes competitive with SOTA multilevel methods while retaining runtime linear in the total hyperedge degree. Its performance derives from efficient spectral-embedding construction, end-to-end projection optimization in the embedding space, and a fully GPU-accelerated implementation. Future directions include extending the framework to support non-uniform vertex and hyperedge weights and integrating HySpecPro with optimization engines in the EDA flow.



\bibliographystyle{ACM-Reference-Format}
\bibliography{ref}

@inproceedings{karypis1997multilevel,
  title={Multilevel hypergraph partitioning: Application in VLSI domain},
  author={Karypis, George and Aggarwal, Rajat and Kumar, Vipin and Shekhar, Shashi},
  booktitle={Proceedings of Design Automation Conference (DAC)},
  pages={526--529},
  year={1997}
}

@inproceedings{murray2013titan,
  title={Titan: Enabling large and complex benchmarks in academic CAD},
  author={Murray, Kevin E and Whitty, Scott and Liu, Suya and Luu, Jason and Betz, Vaughn},
  booktitle={Proceedings of International Conference on Field programmable Logic and Applications (FPL)},
  pages={1--8},
  year={2013}
}

@inproceedings{schlag2016k,
  title={K-way hypergraph partitioning via n-level recursive bisection},
  author={Schlag, Sebastian and Henne, Vitali and Heuer, Tobias and Meyerhenke, Henning and Sanders, Peter and Schulz, Christian},
  booktitle={Proceedings of Workshop on Algorithm Engineering and Experiments (ALENEX)},
  pages={53--67},
  year={2016}
}

@misc{hansen2019pycma,
  author       = {Nikolaus Hansen and Youhei Akimoto and Petr Baudis},
  title        = {{CMA-ES/pycma} on {G}ithub},
  month        = feb,
  year         = 2019,
  url          = {https://doi.org/10.5281/zenodo.2559634},
}

@article{nishino2017cupy,
  title={Cupy: A numpy-compatible library for nvidia gpu calculations},
  author={Nishino, Royud and Loomis, Shohei Hido Crissman},
  journal={Proceedings of Confernce on Neural Information Processing Systems (NeurIPS)},
  volume={151},
  number={7},
  year={2017}
}

@inproceedings{liang2024medpart,
  title={Medpart: A multi-level evolutionary differentiable hypergraph partitioner},
  author={Liang, Rongjian and Agnesina, Anthony and Ren, Haoxing},
  booktitle={Proceedings of International Symposium on Physical Design (ISPD)},
  pages={3--11},
  year={2024}
}

@incollection{fender2022rapids,
  title={Rapids cugraph},
  author={Fender, Alex and Rees, Brad and Eaton, Joe},
  booktitle={Massive Graph Analytics},
  pages={483--493},
  year={2022},
  publisher={Chapman and Hall/CRC}
}

@inproceedings{wang2019deep,
  title={Deep graph library: Towards efficient and scalable deep learning on graphs},
  author={Wang, Minjie Yu},
  booktitle={Proceedings of International Conference on Learning Representations Workshop on Representation Learning on Graphs and Manifolds (ICLR)},
  year={2019}
}

@article{karypis1998hmetis,
  title={hMETIS 1.5: A hypergraph partitioning package},
  author={Karypis, George},
  journal={http://www.cs.umn.edu/metis},
  year={1998}
}

@article{gottesburen2024scalable,
  title={Scalable high-quality hypergraph partitioning},
  author={Gottesb{\"u}ren, Lars and Heuer, Tobias and Maas, Nikolai and Sanders, Peter and Schlag, Sebastian},
  journal={ACM Transactions on Algorithms (TALG)},
  volume={20},
  number={1},
  pages={1--54},
  year={2024}
}

@inproceedings{lee2025hyperg,
  title={HyperG: Multilevel GPU-Accelerated k-way hypergraph partitioner},
  author={Lee, Wan Luan and Lin, Dian-Lun and Chiu, Cheng-Hsiang and Schlichtmann, Ulf and Huang, Tsung-Wei},
  booktitle={Proceedings of Asia and South Pacific Design Automation Conference (ASP-DAC)},
  pages={1031--1040},
  year={2025}
}

@article{chan2018spectral,
  title={Spectral properties of hypergraph laplacian and Titaaapproximation algorithms},
  author={Chan, T-H Hubert and Louis, Anand and Tang, Zhihao Gavin and Zhang, Chenzi},
  journal={Journal of the ACM (JACM)},
  volume={65},
  number={3},
  pages={1--48},
  year={2018}
}

@inproceedings{bustany2023open,
  title={An open-source constraints-driven general partitioning multi-tool for VLSI physical design},
  author={Bustany, Ismail and Gasparyan, Grigor and Kahng, Andrew B and Koutis, Ioannis and Pramanik, Bodhisatta and Wang, Zhiang},
  booktitle={Proceedings of International Conference on Computer Aided Design (ICCAD)},
  pages={1--9},
  year={2023},
  organization={IEEE}
}

@misc{triton,
  author       = {Bustany, Ismail and Gasparyan, Grigor and Kahng, Andrew B and Koutis, Ioannis and Pramanik, Bodhisatta and Wang, Zhiang},
  title        = {TritonPart Official Implementation},
  month        = feb,
  year         = 2023,
  url          = {https://github.com/ABKGroup/TritonPart},
}

@article{hartmanis1982computers,
  title={Computers and intractability: a guide to the theory of np-completeness},
  author={Hartmanis, Juris},
  journal={Siam Review (SIREV)},
  volume={24},
  number={1},
  pages={90},
  year={1982}
}

@article{sajadinia2025shypar,
  title={SHyPar: A Spectral Coarsening Approach to Hypergraph Partitioning},
  author={Sajadinia, Hamed and Aghdaei, Ali and Feng, Zhuo},
  journal={IEEE Transactions on Computer-Aided Design of Integrated Circuits and Systems (TCAD)},
  year={2025}
}

@inproceedings{loukas2018spectrally,
  title={Spectrally Approximating Large Graphs with Smaller Graphs},
  author={Loukas, Andreas and Vandergheynst, Pierre},
  booktitle={Proceedings of International Conference on Machine Learning (ICML)},
  pages={3243--3252},
  year={2018}
}

@inproceedings{zhuo:dac18,
  title={Similarity-Aware Spectral Sparsification by Edge Filtering},
  author={Feng, Zhuo},
  booktitle={Proceedings of Design Automation Conference (DAC)},
  pages={1--6},
  year={2018},
}

@article{teng2016scalable,
  title="{Scalable Algorithms for Data and Network Analysis}",
  author={Teng, Shang-Hua},
  journal={Foundations and Trends in Theoretical Computer Science (FnT)},
  volume={12},
  number={1--2},
  pages={1--274},
  year={2016},
  publisher={Now Publishers, Inc.}
}

@inproceedings{xueqian:iccad14,
  title="{An Efficient Spectral Graph Sparsification Approach to Scalable Reduction of Large Flip-chip Power Grids}",
  author={Zhao, Xueqian and Feng, Zhuo and Zhuo, Cheng},
  booktitle={Proceedings of Internation Conference on Computer-Aided Design (ICCAD)},
  pages={218--223},
  year={2014},
}

@inproceedings{bustany2022specpart,
  title={SpecPart: A supervised spectral framework for hypergraph partitioning solution improvement},
  author={Bustany, Ismail and Kahng, Andrew B and Koutis, Ioannis and Pramanik, Bodhisatta and Wang, Zhiang},
  booktitle={Proceedings of International Conference on Computer-Aided Design (ICCAD)},
  pages={1--9},
  year={2022}
}

@article{bustany2023k,
  title={K-SpecPart: Supervised embedding algorithms and cut overlay for improved hypergraph partitioning},
  author={Bustany, Ismail and Kahng, Andrew B and Koutis, Ioannis and Pramanik, Bodhisatta and Wang, Zhiang},
  journal={IEEE Transactions on Computer-Aided Design of Integrated Circuits and Systems (TCAD)},
  volume={43},
  number={4},
  pages={1232--1245},
  year={2023}
}

@article{chan2020generalizing,
  title={Generalizing the hypergraph laplacian via a diffusion process with mediators},
  author={Chan, T-H Hubert and Liang, Zhibin},
  journal={Theoretical Computer Science (TCS)},
  volume={806},
  pages={416--428},
  year={2020},
  publisher={Elsevier}
}

@article{hmetis,
  title={Multilevel k-way hypergraph partitioning},
  author={Karypis, George and Kumar, Vipin},
  journal={VLSI design},
  volume={11},
  number={3},
  pages={285--300},
  year={2000},
  publisher={Hindawi}
}

@article{spielman2011spectral,
  title={Spectral sparsification of graphs},
  author={Spielman, Daniel and Teng, ShangHua},
  journal={SIAM Journal on Computing (SIAM J. Comput.)},
  volume={40},
  number={4},
  pages={981--1025},
  year={2011},
  publisher={SIAM}
}

@article{spielman2014sdd,
  title="{Nearly Linear Time Algorithms for Preconditioning and Solving Symmetric, Diagonally Dominant Linear Systems}",
  author={Spielman, D. and Teng, S.},
  journal={SIAM Journal on Matrix Analysis and Applications (SIAM J. Matrix Anal. Appl.)},
  volume={35},
  number={3},
  pages={835--885},
  year={2014},
}

\end{document}